\documentstyle[aps,prd,preprint,tighten]{revtex}
\newcommand{\half}{{\textstyle{1\over2}}}

\newcommand{\sgn}{\,\mathrm{sgn}\,}

\begin{document}
\preprint{\vtop{\hbox{MCTP-01-48}\hbox{hep-th/0110213}\vskip24pt}}

\title{Exact multi-membrane solutions in $AdS_7$}

\author{James T.~Liu and W.Y.~Wen\footnote{Email addresses:
\{jimliu, wenw\}@umich.edu}}

\address{Michigan Center for Theoretical Physics\\
Randall Laboratory, Department of Physics, University of Michigan\\
Ann Arbor, MI 48109--1120, USA}

\maketitle

\begin{abstract}
We study the properties of an exact multi-membrane solution in
seven-dimensional maximal $SO(5)$-gauged supergravity.  Unlike previously
known multi-centered solutions, the present one is asymptotically
anti-de Sitter.  We show that this multi-membrane configuration preserves
only a quarter of the supersymmetries.  When lifted to eleven dimensions,
this solution is interpreted as a set of open membranes ending on self-dual
strings on a stack of M5-branes, in the near M5 limit.
\end{abstract}


\ifpreprintsty\else
\begin{multicols}{2}
\fi
\narrowtext

%%%%%%%%%%%%%%%%%%%%%%%%%%%%%%%%%%%%%%%%%%%%%%%%%%%%%%%%%%%%%%%%%%%%%%%%%%%%%%
%% Introduction

%%%%%%%%%%%%%%%%%%%%
\section{Introduction}
%%%%%%%%%%%%%%%%%%%%

The investigation of $p$-brane solutions in supergravities has been an
important aspect in the study of M-theory.  Understanding the basic
properties of $p$-branes has led to important ideas such as dualities,
D-branes, and the AdS/CFT conjecture.  By now, general $p$-brane solutions
in ungauged supergravities are well understood, and much progress has been
made on their classification.  However much less is know about a perhaps
equally important set of objects, namely $p$-branes in anti-de Sitter space,
viewed as solutions of corresponding gauged supergravities.  A main reason
behind this lack of development is that flatness and Ricci flatness has
generally been an integral part of the construction of such solutions.  For
branes that are asymptotically anti-de Sitter, one no longer has the
simplifying condition of Ricci flatness.  Along the same lines, it has been
argued that {\it static} extended objects in anti-de Sitter space cannot
satisfy a no-force condition, as the background curvature would otherwise
provide an unbalanced cosmological force.  These complications of having
curved world-sheets and possible time dependence seem inherent in the study
of $p$-branes in gauged supergravities.

Of course, many of the above difficulties are avoided for the case of
$0$-branes, or black holes.  In fact both de Sitter and anti-de Sitter
black holes have been known for a long time.  More recently, black hole
solutions to gauged supergravities have been constructed and studied in
four \cite{romans,cdk,dk,was}, five \cite{bcs,bcvs} and seven
\cite{cg,lm} dimensions.  However all such constructions have focused
on single centered (although possibly multiply charged) black holes.
Finding multi-centered solutions has generally been as elusive as finding
extended object solutions for much the same reasons.%
\footnote{Partial progress was made in Ref.~\cite{ls} which
constructed multi-centered Euclidean AdS$_5$ black holes.}

For the case of extended $p$-branes, an initial attempt at constructing
a membrane solution to maximal gauged supergravity in seven dimensions
was made in Ref.~\cite{lm}.  While no explicit solution was given, important
properties such as the nature of charges and supersymmetry were
investigated.  An exact magnetic string solution to $N=2$ gauged
supergravity in five dimensions was also constructed and investigated in
Refs.~\cite{Chamseddine:2000xk,Klemm:2000nj}.
In a later development, L\"u and Pope demonstrated a
braneworld reduction of gauged supergravity where the bulk theory reduces
to an ungauged theory on the brane preserving half of the original
supersymmetries \cite{lupope1,lupope2}.  This L\"u-Pope ansatz provided a
breakthrough in the construction of $p$-branes in gauged supergravities
by providing a means of lifting well-known solutions of the
lower-dimensional ungauged theory to yield new solutions to the gauged
supergravity theory.

Consistency of the L\"u-Pope ansatz provides serious restrictions on the
fields that may be lifted or reduced.  For bosonic fields, in addition to
the graviton, only scalars and $n$-form potentials satisfying
odd-dimensional self-duality equations \cite{tpvn,ppvn} may be
consistently reduced, yielding scalars and $n$-form field strengths,
respectively, on the brane.  Hence only $p$-branes charged under the
appropriate fields may be constructed through this lifting technique.
Furthermore, use of odd-dimensional self-dual fields suggests that the
resulting $p$-branes must necessarily be dyonically charged.  Nevertheless,
the L\"u-Pope ansatz opens up the possibility of constructing new solutions
to gauged supergravities.

In this letter, we revisit the seven-dimensional system and find an exact
multi-membrane solution by applying the L\"u-Pope ansatz to the self-dual
string in six dimensions.  Examination of supersymmetry indicates that
only a quarter of the original supersymmetries survive.  This is in contrast
to the basic membrane solution of ungauged supergravity, which preserves half
of the supersymmetries.

When lifted to eleven dimensions, this multi-membrane solution
corresponds to the near M5-brane horizon limit of a set of open
membranes ending on self-dual strings on a stack of M5-branes
\cite{Strominger:1996ac,Townsend:1996af}.  Since the self-dual strings
of the $(2,0)$ theory preserve only eight supercharges, this explains
why no supersymmetry restoration takes place in the present configuration.

%%%%%%%%%%%%%%%%%%%%
\section{$N=4$, $D=7$ gauged supergravity}
%%%%%%%%%%%%%%%%%%%%

The bosonic sector of maximal $SO(5)$ gauged supergravity in seven dimensions
contains a graviton, ten Yang-Mills fields transforming in the adjoint of
$SO(5)_g$, five 3-forms in the $\bf5$ of $SO(5)_g$ and fourteen scalars which
parametrize the coset $SL(5,R)/SO(5)_c$.  In addition, there are four
gravitini and sixteen spin-${1\over2}$ fields transforming as the $\bf4$ and
$\bf16$ of $SO(5)_c$ respectively.  The bosonic Lagrangian is given by
\cite{ppvn,ppvnw}:
\begin{eqnarray}
\label{action}
e^{-1}{\cal L} &=&
R + \half g^2 (T^2 - 2T_{ij}T^{ij}) - Tr(P_{\hat\mu} P^{\hat\mu})
-\half (V_I{}^i V_J{}^j F_{\hat\mu\hat\nu}^{IJ})^2
-{\textstyle{1\over12}} (V_i^{-1\,I} S_{\hat\mu\hat\nu\hat\rho}^I)^2\nonumber\\
&&+ e^{-1}g^{-1}\Bigl[{\textstyle{1\over2}} S_{[3]}^I\wedge dS_{[3]}^I
+{\textstyle{1\over2}}
\epsilon_{IJKLM}S_{[3]}^I\wedge F_{[2]}^{JK}\wedge F_{[2]}^{LM}
+ \Omega_7(A, F) \Bigr].
\end{eqnarray}
Here we follow the notation of \cite{lupope1,lupope2}.  In particular, we
work with a metric of signature $(-+\ldots+)$ and the 3-forms $S_{[3]}^I$ 
are rescaled from those given in Ref.~\cite{ppvn}.  Upper case indices
$I,J = 1, \ldots, 5$ denote $SO(5)_g$ indices, while lower case $i,
j = 1, \ldots, 5$ denote $SO(5)_c$ indices.  Finally, $\Omega_7$ is a
Chern-Simons form built from the gauge fields.  As we will set
$A_{[1]}^{IJ}=0$, we will not need its explicit form.

The fourteen scalar degrees of freedom are contained in the coset elements
$V_I{}^i$,
transforming as a $5$ under both $SO(5)_g$ and $SO(5)_c$.  These funfbeins
may then be used to build the $T$-tensor, $T_{ij} = V_i^{-1\,I}
V_j^{-1\,I}$, with trace $T\equiv T_{ij} \delta^{ij}$.  The scalar kinetic
term, $P_{\hat\mu}^{ij}$, and composite $SO(5)_c$ connection, $Q_\mu{}^i{}_j$,
are defined through $V_i^{-1\,I}{\cal D}_{\hat\mu}V_I{}^j=(Q_{\hat\mu})_{[ij]}
+(P_{\hat\mu})_{(ij)}$, where ${\cal D}_{\hat\mu}$ is the gauge covariant
derivative.

The relevant supersymmetries are given by
\begin{eqnarray}
\label{grsusy}
\delta\psi_{\hat\mu}&=&\Bigl[{\cal D}_{\hat\mu}+{g\over20}T\gamma_{\hat\mu}
-{1\over40}(\gamma_{\hat\mu}{}^{\hat\nu\hat\lambda}-8\delta_{\hat\mu}^{\hat\nu}
\gamma^{\hat\lambda}) \Gamma^{ij}V_I{}^iV_J{}^jF_{\hat\nu\hat\lambda}^{IJ}
\nonumber\\
&&+{1\over60}(\gamma_{\hat\mu}{}^{\hat\nu\hat\lambda\hat\sigma}
-{\textstyle{9\over2}}\delta_{\hat\mu}^{\hat\nu}
\gamma^{\hat\lambda\hat\sigma})\Gamma^iV_i^{-1\,I}
S_{\hat\nu\hat\lambda\hat\sigma}^I \Bigr]\epsilon,\\
\label{disusy}
\delta\lambda_i&=&\Bigl[{g\over2}(T_{ij}-{\textstyle{1\over5}}
\delta_{ij}T)\Gamma^j
+{1\over2}\gamma^{\hat\mu} P_{\hat\mu\,ij}\Gamma^j
+{1\over16}\gamma^{\hat\mu\hat\nu}(\Gamma^{kl}\Gamma^i-{\textstyle{1\over5}}
\Gamma^i \Gamma^{kl})V_K{}^kV_L{}^lF_{\hat\mu\hat\nu}^{KL}\nonumber\\
&&+{1\over120}\gamma^{\hat\mu\hat\nu\hat\lambda}(\Gamma^{ij}-4\delta^{ij})
V_j^{-1\,J}S_{\hat\mu\hat\nu\hat\lambda}^J\Bigr]\epsilon.
\end{eqnarray}
Note that consistency requires both $\Gamma^i\lambda_i=0$ and
$\Gamma^i\delta\lambda_i=0$.

In Refs.~\cite{lupope1,lupope2}, it was argued that this maximal gauged
$N=4$ theory in seven dimensions may be reduced to yield ungauged $N=(2,0)$
pure supergravity in six dimensions.  We recall that the latter theory
contains a $D=6$ graviton, five 2-forms with self-dual field strength
[transforming as a vector of {\it global} $SO(5)$] and four gravitini, but
no scalars nor Yang-Mills fields.  This suggests that an appropriate
braneworld reduction is simply to take trivial scalars, $V_I^i=\delta_I^i$,
and vanishing gauge fields, $A_{\hat\mu}^{IJ}=0$.  The consistency of this
truncation was demonstrated in Ref.~\cite{lupope2}, provided
\begin{equation}
S_{[3]}^I\wedge S_{[3]}^J=0,\qquad *S_{[3]}^I\wedge S_{[3]}^J=0,
\end{equation}
for all $I, J$.  In this case, the Lagrangian (\ref{action}) simplifies to
\begin{equation}
\label{act2}
e^{-1}{\cal L}=R+{\textstyle{15\over2}}g^2-{\textstyle{1\over12}}
(S_{\hat\mu\hat\nu\hat\rho}^I)^2+e^{-1}\left[{\textstyle{1\over2}}g^{-1}
S_{[3]}^I\wedge dS_{[3]}^I\right],
\end{equation}
while the supersymmetry transformations become
\begin{eqnarray}
\label{grs2}
\delta\psi_{\hat\mu}&=&\Bigl[\nabla_{\hat\mu}+{g\over4}\gamma_{\hat\mu}
+{1\over60}(\gamma_{\hat\mu}{}^{\hat\nu\hat\lambda\hat\sigma}
-{\textstyle{9\over2}}\delta_{\hat\mu}^{\hat\nu}
\gamma^{\hat\lambda\hat\sigma})\Gamma^IS_{\hat\nu\hat\lambda\hat\sigma}^I
\Bigr]\epsilon,\\
\label{dis2}
\delta\lambda_I&=&\Bigl[
{1\over120}\gamma^{\hat\mu\hat\nu\hat\lambda}(\Gamma^{IJ}-4\delta^{IJ})
S_{\hat\mu\hat\nu\hat\lambda}^J\Bigr]\epsilon.
\end{eqnarray}
%

%%%%%%%%%%%%%%%%%%%%
\section{The multi-membrane solution}
%%%%%%%%%%%%%%%%%%%%

The L\"u-Pope ansatz provides a consistent means of
relating gauged supergravities to ungauged counterparts in one lower
dimension with half of the original supersymmetries preserved.  In the context
of the Randall-Sundrum braneworld, the reduction ansatz for the above
seven-dimensional theory is
\cite{lupope1,lupope2}:
\begin{eqnarray}
ds_7^2&=&e^{-2k|z|}g_{\mu\nu}(x)dx^\mu dx^\nu+dz^2,\nonumber\\
S_{[3]}^I&=&e^{-2k|z|}H_{[3]}^I(x).
\end{eqnarray}
As indicated in \cite{Alonso-Alberca:2000ne,Falkowski:2000er,bkvp},
preservation of supersymmetry on the brane
demands that $g$ changes sign when passing through the brane,
{\it i.e.}~$g=2k\sgn(z)$.

While presented above as a braneworld reduction, it is important to note
that the L\"u-Pope ansatz is equally valid as a consistent reduction of
the bulk gauged supergravity to the ungauged $N=(2,0)$ theory
\cite{lupope1}.  For the supergravity reduction without the brane, one
simply removes the absolute values and takes $g=2k$.  The reduction
ansatz may then be viewed as a reduction of AdS$_7$ along horospherical
slices.  This is the primary point of view we take in the present case.%
\footnote{Nevertheless we maintain the absolute values in appropriate
equations, since once removed they are harder to reinsert.}

Substituting this ansatz into the equations of motion resulting from
(\ref{act2}) yields the bosonic equations of the six-dimensional $N=(2,0)$
theory:
\begin{eqnarray}
&\displaystyle dH_{[3]}^I=0,\qquad H_{[3]}^I=*_6 H_{[3]}^I,&\nonumber\\
&\displaystyle R_{\mu\nu}={\textstyle{1\over4}}H_{\mu\rho\sigma}^I
H_\nu^{I\,\rho\sigma}.
\end{eqnarray}
While in principle the six-dimensional gravitino variation
\begin{equation}
\delta\psi_\mu=[\nabla_\mu+{\textstyle{1\over4}}H_{\mu\nu\rho}^I
\gamma^{\nu\rho} \Gamma^I]\epsilon,
\end{equation}
may be obtained from the reduction of (\ref{grs2}) following the procedure of
Ref.~\cite{dls}, we will instead work directly with the seven dimensional
transformations (\ref{grs2}) and (\ref{dis2}).

In general, six-dimensional supergravities support the construction of BPS
dyonic string solutions.  However only the self-dual string is of relevance
in the $N=(2,0)$ case.  By global $SO(5)$ invariance, the self-dual string
solution may be rotated so that it carries charge under a single 2-form
potential.  Specifying this $SO(5)$ direction by a unit vector $\hat n^I$,
the multi-string solution has the form \cite{dlblack,dfkr,lupope1}
\begin{eqnarray}
&&ds_6^2={\cal H}^{-1}(y)\eta_{\mu\nu}dx^\mu dx^\nu
+{\cal H}(y)d\vec y\,^2\nonumber\\
&&H_{[3]}^I=dB_{[2]}^I+*_6dB_{[2]}^I,\qquad B_{01}^I={\cal H}^{-1}(y)\hat n^I.
\end{eqnarray}
The $x^\mu$, $\mu=0,1$ are coordinates longitudinal to the string, while the
$y^i$, $i=2,3,4,5$ are transverse coordinates.
The function ${\cal H}(y)$ is harmonic in the transverse space.  For a flat
transverse space, it may take on the simple multi-string form
\begin{equation}
{\cal H}(y)=1+\sum_{i=1}^{m} {q_i\over|\vec y-\vec y_i|^2},
\end{equation}
where $\{q_i\}$ and $\{\vec y_i\}$ denote respectively the charges and
centers of the strings.

This solution may be lifted to seven dimensions, yielding \cite{lupope1}
\begin{eqnarray}
\label{eq:mmsol}
&&ds_7^2=e^{-2k|z|}\left[{\cal H}^{-1}(y)\eta_{\mu\nu}dx^\mu dx^\nu
+{\cal H}(y)d\vec y\,^2\right]+dz^2,\nonumber\\
&&S_{[3]}^I=e^{-2k|z|}[dB_{[2]}^I+*_6dB_{[2]}^I],
\qquad B_{01}^I={\cal H}^{-1}(y)\hat n^I.
\end{eqnarray}
As a braneworld configuration, this may be interpreted as a multi-string
configuration on the 5-brane.  Viewed from the bulk, on the other hand,
this solution corresponds to multi-membranes ending on self-dual strings
on the 5-brane, where coordinates on the membrane are $x^0$, $x^1$, and $z$.
This latter interpretation is obtained by removing the absolute values, thus
pushing the Randall-Sundrum brane off to the boundary of AdS$_7$ at
$z\to-\infty$.

%%%%%%%%%%%%%%%%%%%%
\section{Supersymmetry}
%%%%%%%%%%%%%%%%%%%%

We examine the supersymmetry directly in seven dimensions.  Starting from
the spin-$\half$ variation, (\ref{dis2}), we first compute
\begin{equation}
\gamma\cdot S^I = -6e^{k|z|}{\cal H}^{-3/2}\gamma^{\overline{01i}}
\partial_i{\cal H} \hat n^I(1+\gamma^7),
\end{equation}
to obtain
\begin{equation}
\label{eq:lamsusy}
\delta\lambda_I=-{\textstyle{1\over20}}e^{k|z|}{\cal H}^{-3/2}
(\Gamma^I\hat n\cdot\Gamma-5\hat n^I)\gamma^{\overline{01i}}
\partial_i{\cal H}(1+\gamma^7)\epsilon.
\end{equation}
Note that $\gamma^7$ is the six-dimensional chirality operator,
$\gamma^7=\gamma^{\overline{0}}\gamma^{\overline{1}}\cdots
\gamma^{\overline{5}}$, which may be identified with the additional
Dirac matrix in seven dimensions.  As will be seen below, we find it
convenient to take $\gamma^{\overline{z}}=-\gamma^7$.  In any case,
preservation of supersymmetry demands that the Killing spinors vanish
under the projection $P_+\epsilon=0$ where
\begin{equation}
P_{\pm}=\half(1\pm\gamma^7).
\end{equation}
This requirement on Killing spinors holds independent of the absolute value
$|z|$ in the solution, and hence is unrelated to the presence of a `kinked'
Randall-Sundrum brane.  In fact, this result follows simply from
satisfying odd-dimensional self-duality, as was already noted in \cite{lm}.

For the gravitino, we consider the three separate variations,
$\delta\psi_\mu$, $\delta\psi_i$ and $\delta\psi_z$.  After considerable
manipulation of (\ref{grs2}), we obtain
\begin{eqnarray}
\label{eq:psisusy}
\delta\psi_\mu&=&[\partial_\mu+{g\over4}\gamma_\mu(1-{2k\sgn z\over
g}\gamma^z)-{1\over2}e^{k|z|}{\cal H}^{-3/2}\gamma^{\overline{i}}
\partial_i{\cal H}(\widetilde P_-
+{2\over5}\gamma^{\overline{01}}\hat n\cdot\Gamma P_+)]\epsilon\nonumber\\
\delta\psi_i&=&[\partial_i+{1\over4}{\cal H}^{-1}\partial_i{\cal H}
+{g\over4}\gamma_i(1-{2k\sgn z\over g}\gamma^z)\nonumber\\
&&\qquad\qquad
-{1\over2}\gamma^j\gamma_i{\cal H}^{-1}\partial_j{\cal H}\widetilde P_-
-{1\over10}\hat n\cdot\Gamma {\cal H}^{-1}\partial_j{\cal H}
(2\delta_i^j-3\gamma_i{}^j)\gamma^{\overline{01}}P_+]\epsilon\nonumber\\
\delta\psi_z&=&[\partial_z+{g\over4}\gamma_z
-{1\over5}\gamma_z\gamma^{\overline{01i}}e^{k|z|}{\cal H}^{-3/2}
\partial_i{\cal H} \hat n\cdot\Gamma P_+]\epsilon
\end{eqnarray}
where the second projection $\widetilde P_\pm$ is given by
\begin{equation}
\widetilde P_\pm=\half(1\pm\gamma^{\overline{01}}\hat n\cdot\Gamma).
\end{equation}
Note that $\widetilde P_\pm$ commutes with $P_\pm$ and is suggestive of an
electric BPS string lying in the 0--1 directions.

Examination of (\ref{eq:psisusy}) verifies the consistency of the choices
$g=2k\sgn z$ and $\gamma^z=-\gamma^7$.  The Killing spinor equations are
then solved by
\begin{equation}
\label{eq:mmks}
\epsilon=e^{-{k\over2}|z|}{\cal H}^{-1/4}P_-\widetilde P_+\epsilon_0,
\end{equation}
where $\epsilon_0$ is a constant spinor.  This demonstrates that the
multi-membrane solutions preserve only a quarter of the maximum
supersymmetries.

%%%%%%%%%%%%%%%%%%%%
\section{The non-extremal membrane}
%%%%%%%%%%%%%%%%%%%%

While our focus has been on multi-centered BPS configurations, we note
that the L\"u-Pope ansatz equally well allows the lifting of a black
string solution to AdS$_7$.  Blackening of the self-dual string is
straightforward \cite{Duff:1996hp}, and the resulting black membrane
solution may be written as
\begin{eqnarray}
\label{eq:mblack}
&&ds_7^2=e^{-2k|z|}\left[{\cal H}^{-1}(-fdt^2+dx^2)
+{\cal H}\Bigl({dr^2\over f}+r^2d\Omega_3^2\Bigr)\right]+dz^2,\nonumber\\
&&S_{[3]}^I=e^{-2k|z|}[dB_{[2]}^I+*_6dB_{[2]}^I],
\qquad B_{01}^I=\coth\mu\,{\cal H}^{-1}\hat n^I.
\end{eqnarray}
The non-extremality function $f$ and the harmonic function $\cal H$
specifying the solution are given by
\begin{equation}
f=1-{k\over r^2},\qquad{\cal H}=1+{k\sinh^2\mu\over r^2}.
\end{equation}
The resulting self-dual charge is given by $q=k\sinh\mu\cosh\mu$.
The black membrane has a horizon located at $r_+=k^{1/2}$,
with topology $R\times R^+\times S^3$ with the second
factor corresponding to the $z$ direction.  The extremal limit may be
obtained by taking $k\to0$ and $\mu\to\infty$ with $q$ held fixed.

%%%%%%%%%%%%%%%%%%%%
\section{Discussion}
%%%%%%%%%%%%%%%%%%%%

We again emphasize that, while motivated by the search for braneworld
reductions, the above technique for constructing multi-membrane
solutions in anti-de Sitter backgrounds holds equally well in the bulk.
It is instructive to reexamine the form of the solution, (\ref{eq:mmsol}),
where the metric takes on a horospherical form
\begin{equation}
\label{eq:mmads}
ds_7^2=e^{-gz}\left[{\cal H}^{-1}(y)\eta_{\mu\nu}dx^\mu dx^\nu
+{\cal H}(y)d\vec y\,^2\right] +dz^2.
\end{equation}
It is evident that this solution is asymptotically anti-de Sitter far
from the membranes when ${\cal H}\to 1$.  In addition, however, the
metric longitudinal to the membrane has the form
\begin{equation}
ds^2=e^{-gz}{\cal H}^{-1}(-dt^2+dx^2)+dz^2,
\end{equation}
which asymptotes to AdS$_3$.  Thus, unlike their flat space counterparts,
membranes in anti-de Sitter space necessarily have curved world-sheets.

It should be mentioned that the metric ansatz leading to
(\ref{eq:mmads}) was considered in Ref.~\cite{lm} without success.
However in that case, the difficulty was presumably in attempting to seek
a half-BPS configuration.  As demonstrated above, the multi-membrane
solution here necessarily preserves only a quarter of the
supersymmetries, as indicated in (\ref{eq:mmks}).  Since AdS$_7$ is
maximally symmetric and maximally supersymmetric, this quarter-BPS
feature is perhaps somewhat unusual.  Examining the two simultaneous
broken supersymmetries,%
\footnote{This presence of two complimentary supersymmetry projections
was already anticipated in Ref.~\cite{lm}.}
corresponding to projections $P_+$ and
$\widetilde P_-$, we note that the former selects a definite
$\gamma^{\overline z}$ eigenvalue.  Thus, for Killing spinors satisfying
both projections, the latter $\widetilde P_\pm$ may be replaced by the
equivalent
\begin{equation}
\widetilde{\kern-2pt\widetilde P}_\pm=\half(1\pm\gamma^{\overline{01z}}
\hat n\cdot\Gamma),
\end{equation}
which has a conventional form for an electrically charged membrane.

In this sense, the new feature of the membrane solution to gauged
supergravity is the projection $P_\pm$ related to both odd-dimensional
self duality and AdS Killing spinors.  We recall that, for pure AdS in
horospherical coordinates [{\it i.e.}~for ${\cal H}=1$ in (\ref{eq:mmads})],
the Killing spinors take the form
\begin{equation}
\epsilon_-=e^{-gz/4}P_-\epsilon^0,\qquad
\epsilon_+=e^{gz/4}[1-{\textstyle{g\over2}}(x^\mu\gamma_\mu+y^i\gamma_i)]
P_+\epsilon^0
\end{equation}
While $\epsilon_-$ is straightforward, we see that $\epsilon_+$ is
sensitive to the horizon ($z\to\infty$), and furthermore has non-trivial
dependence on all coordinates.  Thus the latter Killing spinor cannot
survive once the membrane solution is turned on.  Viewing AdS$_7$ as
the near-horizon limit of a stack of M5-branes, we recall that
odd-dimensional self-duality for $S_{[3]}^I$ originates from the action
of $F_{[4]}$ in eleven dimensions.  The latter, and in particular the
Chern-Simons term $F_{[4]}\wedge F_{[4]}\wedge A_{[3]}$, plays an
important role in determining possible intersecting brane solutions of
M-theory.  Since the multi-membrane solution lifts to a configuration of
open M2-branes ending on a stack of M5-branes, the projection $P_\pm$ may
be identified with that of a M5-brane, while the projection
$\widetilde{\kern-2pt\widetilde P}_\pm$ with that of a M2-brane.
In the absence of a M2-brane, supersymmetry is enhanced in the near M5
limit, yielding the additional Killing spinors $\epsilon_+$ above.
However, for the complete solution, the nature of the intersection
prevents any enhancement of supersymmetry.

While the L\"u-Pope ansatz sets the $SO(5)$ gauge fields to zero, it is
curious to note that the multi-membrane solution preserves $SO(4)\subset
SO(5)$.  This hints that it may be possible to turn on $SO(4)$ gauge
fields orthogonal to the direction given by $\hat n^I$ without breaking
any further supersymmetries \cite{lm}.  From a braneworld perspective,
it is unclear what effect bulk gauge fields may have, and in particular
it is expected that such fields cannot localize on the brane.  Thus there
may be possible braneworld implications for the localization of bulk
Yang-Mills fields if any membrane configurations with further non-trivial
breakings of the R-symmetry may be constructed.

Use of the L\"u-Pope ansatz restricts consideration to a Poincar\'e
patch of AdS$_7$.  It remains an open issue whether the multi-membrane
solution (\ref{eq:mmsol}) may be transformed to some form of global AdS
coordinates, or whether the singularity and horizon structure prevents
this.  Since one has the geometry of open membranes ending on self-dual
strings, this issue is analogous to that of explicit construction of
strings ending on the boundary of AdS, as considered in the AdS$_5$
case of the AdS/CFT duality.  Curiously, in this AdS$_5$ case, a
string solution was obtained in Refs.~\cite{Chamseddine:2000xk,Klemm:2000nj}
using cylindrically symmetric coordinates.  This string is magnetically
charged under the $N=2$ graviphoton and (similar to the membranes
discussed here) preserves a quarter of the supersymmetries.

Finally, understanding open membranes ending on the fivebrane may
be important in developing a better picture of the $N=(2,0)$ theory
underlying the dynamics of the M5-brane.  While much work has been focused
on the world-sheet point of view, soliton solutions like the one presented
here allow for a complimentary approach to such investigations.  It is
natural to suspect that gravity may be decoupled from the system, in
which case one expects to obtain results consistent with that of open
membrane theory \cite{Gopakumar:2000ep,Bergshoeff:2000ai}.  In fact,
this connection to open membranes was already shown to occur in the near
horizon limit of the self-dual string in the M5-brane world-sheet approach
\cite{Berman:2001fs}.  It would be interesting to carry out a similar
analysis for the multi-membrane solution given above.

%%%%%%%%%%%%%%%%%%%%%%%%%%%%%%%%%%%%%%%%%%%%%%%%%%%%%%%%%%%%%%%%%%%%%%%%%%%%%%
\section*{Acknowledgments}
This research was supported in part by DOE Grant DE-FG02-95ER40899 Task G.

%%%%%%%%%%%%%%%%%%%%%%%%%%%%%%%%%%%%%%%%%%%%%%%%%%%%%%%%%%%%%%%%%%%%%%%%%%%%%%

\ifpreprintsty\else
\end{multicols}
\fi

%%%%%%%%%%%%%%%%%%%%%%%%%%%%%%%%%%%%%%%%%%%%%%%%%%%%%%%%
\end{document}